# Policies frozen in silicon: using WPR to expose the politics of problem-solution configurations in technical artifacts

Jörgen Behrendtz

Dept. of Media Studies, Stockholm University, jorgen.behrendtz@ims.su.se

Lina Rahm

Division of History of Science, Technology and Environment, Royal Institute of Technology (KTH), lina.rahm@kth.se

Design is often characterized as an act of problem-solving. This is a perspective that, while pervasive, risks reducing complex socio-technical conditions to easily fixable issues. This paper critiques the ideology of "design as problem-solving", highlighting its culmination in technological solutionism, where societal and human challenges are reframed as technical problems awaiting technical answers. Drawing on critiques and the recognition of "wicked problems", we argue that design must also be understood as a process of problem-framing, emphasizing the interpretive work involved in defining what counts as a problem and why. To advance this analytical perspective, we propose applying the What's the Problem Represented to be? (WPR) approach from critical policy studies to design and technology. By treating artifacts as materialized problem representations, WPR allows for the systematic unpacking of the ideological, cultural, and political assumptions encoded in technological forms. This analytical lens can reveal hidden problematisations within artifacts, foster reflexive design practice, and empirically challenge techno-solutionism. Ultimately, integrating WPR into design research enriches both design theory and philosophy of technology by offering a method to interrogate how technologies shape, and are shaped by, the questions they claim to answer.

## 1 FIRST PREMISE: DESIGN AS PROBLEM-SOLVING

Design is often described as "problem-solving". This is undeniably a generic, yet frequently recurring perspective [1, 2]. It is perhaps not surprising that designers, tasked with ensuring that artifacts are made valuable to all users by addressing aesthetics, usability, ergonomics, safety, marketability, manufacturability, functionality, and sustainability, while responding to market opportunities and needs from corporations, entrepreneurs, consumers, governments, and non-profits, in the end conceptualize of their work as problem-solving [3]. In its most extreme form, the ideology of design as problem-solving can result in an exaggerated version, sometimes referred to as technological solutionism [4, 5]. Here, every societal or human issue is reframed as a "problem" awaiting a technological fix. Urban designer Michael Dobbins [6] refers to the phenomenon this way:

> "The disconnect between problem and solution, always likely to be an issue, became exaggerated in the culture and practice of modernism in city design and planning, where problems were 'dumbed down' to meet the solutions offered." (p.182)

Dobbins effectively underscores the significance of *problem representations* and their constitutive role in shaping what comes to be regarded as an appropriate or "suitable" solution. When developed, the argument suggests that the alignment between problems and solutions, sometimes valorised within design discourse, cannot be understood as a neutral or objective process, but is instead conditioned by psychological, socio-economic, and political factors that influence both how problems are defined and how solutions are imagined. It could even be interpreted as a warning that technological solutionism is deceptive because it is culturally practical, instrumental, and aligns with how technology and industry operate. Thus, technological solutionism is perhaps better seen as a diagnostic case that reveals the risks of treating every design issue as a solvable problem (at least by computer technologies). It is almost as if it mirrors, amplifies and reverses the problem-solving narrative: it assumes that if an acceptable solution can be built, the problem has also been understood and resolved.

## 2 SECOND PREMISE: DESIGN AS PROBLEM-FRAMING

The worries about technological solutionism can be further strengthened by the fact that once a design solution has materialized (e.g. as a media technology), it arguably represents a more rigid, even frozen, proposal for a solution. As Herbert A. Simon [7] puts it:

> "Most design resources go into discovering or generating alternatives, and not into choosing among them. In fact, it is quite common for a single alternative to emerge from the design process—a single plan for a house, or for a bridge, or a single score for a sonata. No choice remains; all of the choosing has been done in the course of generating, selecting among, and combining the elements and components of the design." (p.247)

Naturally, the problem-solving ideology can be, and has also been, criticized for oversimplifying the design *process*[1] [8]. Daly et al. [9] refer to the concept of problem exploration, which they define as "intentionally developing varied perspectives on a problem in order to generate alternative solutions" (p.695). They emphasize that exploring a problem involves more than examining the details of an already defined problem, it also requires investigating the deeper and more diverse meanings of the problem itself. Many design problems are also *wicked* (complex, ill-defined, and inherently political), and design operates in messy, uncertain terrains where there is already conflict over both problem and solution [3, 10, 11]. Seeing how the notion of wicked problems emerged from policy studies, it seems only logical to also return to policy studies for unpacking them as well as the potential tendency towards technological solutionism.

## 3 ARGUMENT: WPR CAN UNPACK WHAT TECHNOLOGIES EMBED

WPR (What's the Problem Represented to be?) is a method stemming from critical political science that addresses problematizations (problem-solution configurations) as ways to expose the possible ideological and political effects of a specific co-configuration of problem and solution [12]. While mainly proposed as a way to examine textual policy, recent studies have developed the approach to technologies, then conceptualized as materialized policies [13, 14, 15].

---

[1] While there is arguably some kind of fundamental truth to the idea that 'solving problems' is at the center of design, we are well aware that this view could also be criticized for oversimplifying the creative, iterative, and contextual nature of design *processes*: it treats challenges as static and objective (when they are systemic and evolving), casts designers as detached experts imposing 'techno fixes' (ignoring collaboration), implies linear processes with optimal outcomes (overlooking exploration and satisficing), and commodifies design for marketing while neglecting broader systemic impacts and stakeholder realities.

Technologies frequently function as broad, adaptive responses to a range of contemporary challenges, yet they simultaneously embody particular, often politically inflected, modes of recording, storing, and transmitting information. This observation invites a series of critical questions: in what ways do the design and implementation of technologies privilege certain social values and interests while marginalising others, and with what consequences for differently positioned groups? How do prevailing narratives of technological neutrality and universality conceal the cultural and political assumptions embedded within such systems? And, finally, what social and experiential realities emerge when complex societal issues are addressed primarily through technology-driven interventions? Designed artifacts embody particular problem representations, and "enclose" what the designer thought the problem was and how it should be addressed. We consequently suggest that WPR:s analytical questions can help unpack what specific designs embed:

- What specific problem(s) does a specific design solution set out to solve?
- What presuppositions or assumptions underlie these representations of the 'problem(s)'?
- How have these representations of the 'problem(s)' come about?
- What is left unproblematic in these problem representations? Where are the silences? Can the 'problem(s)' be thought about differently?
- What effects are produced by these representations of the 'problem(s)'?
- How/where have these representations of the 'problem(s)' been produced, disseminated and defended? How could they be questioned, disrupted, and replaced?

## 4 CONCLUSION

In conclusion, we see several benefits from applying WPR to design, HCI and technological artefacts, 1) It can reveal hidden problematizations in artifacts: by applying WPR to designs (e.g., apps, infrastructures), we can make visible how everyday objects encode assumptions about users, society, and needs, making critique tangible and focused. 2) It may offer a diagnostic toolkit for designers as WPR provides concrete questions to intentionally interrogate one's own designs, fostering reflexive practice around problem-solution configurations. 3) It can challenge techno-solutionism empirically: in case studies of real designs (e.g., surveillance systems, AI for climate change, or smart city tech) WPR can expose limits, like ignored social costs or excluded voices, yielding nuanced alternatives. 4) It can extend WPR's scope: traditionally discourse-focused, WPR gains materiality here, enabling analysis of hybrid socio-technical systems and enriching philosophy of technology with post-structuralist tools.

## REFERENCES

[1] Dorst, K. (2004). On the problem of design problems—problem solving and design expertise. *Journal of Design Research, 4*(2), 185-196.
[2] Foster, M. K. (2021). Design thinking: A creative approach to problem solving. *Management Teaching Review, 6*(2), 123-140.
[3] Boradkar, P. (2010). Design as problem solving. In: *The Oxford handbook of interdisciplinarity (pp. 273-287)*.
[4] Reyes-Cruz, G., Spors, V., Muller, M., Ciolfi Felice, M., Bardzell, S., Williams, R. M., ... & Feldfeber, I. (2025, April). Resisting AI Solutionism: Where Do We Go From Here?. In *Proceedings of the Extended Abstracts of the CHI Conference on Human Factors* in Computing Systems (pp. 1-6).
[5] Sætra, H. S., & Selinger, E. (2024). Technological remedies for social problems: Defining and demarcating techno-fixes and techno-solutionism. *Science and engineering ethics, 30*(6), 60. https://doi.org/10.1007/s11948-024-00524-x
[6] Dobbins, M. (2009). *Urban design and people.* Hoboken: John Wiley & Sons.
[7] Simon, H. A. (1995). Problem forming, problem finding and problem solving in design. In Arne Collen & Wojciech W. Gasparski (Eds.) *Design & systems: General applications of methodology (pp. 245-257).* New Brunswick: Transaction Publishers.
3